\documentclass[10pt,a4paper,onecolumn]{article}

\usepackage[T1]{fontenc}
\usepackage[utf8]{inputenc}

\usepackage{marginnote}
\usepackage{graphicx}
\usepackage{xcolor}
\usepackage{authblk,etoolbox}
\usepackage{titlesec}
\usepackage{calc}
\usepackage{tikz}
\usepackage{hyperref}
\hypersetup{colorlinks, breaklinks=true,
	urlcolor=[rgb]{0.0, 0.5, 1.0},
	linkcolor=[rgb]{0.0, 0.5, 1.0},
	unicode=true,
	pdftitle={The Population Synthesis Toolkit (PST) Python Library},
	pdfborder={0 0 0}}
\usepackage{caption}
\usepackage{tcolorbox}
\usepackage{amssymb,amsmath}
\usepackage{seqsplit}
\usepackage{xstring}
\usepackage{float}
\usepackage[top=3.5cm, bottom=3cm, right=1.5cm, left=1.0cm,
headheight=2.2cm, reversemp, includemp, marginparwidth=4.5cm]{geometry}
\usepackage{fancyhdr}
\usepackage{parskip}

\usepackage[backend=biber]{biblatex}
\bibliography{paper.bib}

\let\origfigure\figure
\let\endorigfigure\endfigure
\renewenvironment{figure}[1][2] {\origfigure[H]} {\endorigfigure}

\let\textttOrig=\texttt
\def\texttt#1{\textttOrig{\seqsplit{#1}}}
\renewcommand{\seqinsert}{\ifmmode\allowbreak\else\penalty6000\hspace{0pt plus 0.02em}\fi}

\makeatletter
\let\href@Orig=\href
\def\href@Urllike#1#2{\href@Orig{#1}{\begingroup\def\Url@String{#2}\Url@FormatString\endgroup}}
\def\href@Notdoi#1#2{\def\tempa{#1}\def\tempb{#2}%
	\ifx\tempa\tempb\relax\href@Urllike{#1}{#2}\else\href@Orig{#1}{#2}\fi}
\def\href#1#2{%
	\IfBeginWith{#1}{https://doi.org}%
	{\href@Urllike{#1}{#2}}{\href@Notdoi{#1}{#2}}}
\makeatother

\newlength{\cslhangindent}
\newlength{\csllabelwidth}
\newenvironment{CSLReferences}[3]%
{%
	\setlength{\parindent}{0pt}
	\ifodd #1 \everypar{\setlength{\hangindent}{\cslhangindent}}\ignorespaces\fi
	\ifnum #2 > 0
	\setlength{\parskip}{#2\baselineskip}
	\fi
}{}

\titleformat{\section}{\normalfont\sffamily\Large\bfseries}{}{0pt}{}
\titleformat{\subsection}{\normalfont\sffamily\large\bfseries}{}{0pt}{}
\titleformat{\subsubsection}{\normalfont\sffamily\bfseries}{}{0pt}{}
\titleformat*{\paragraph}{\sffamily\normalsize}

\makeatletter
\let\ps@plain\ps@fancy
\fancyheadoffset[L]{4.5cm}
\fancyfootoffset[L]{4.5cm}
\makeatother

\providecommand{\tightlist}{\setlength{\itemsep}{0pt}\setlength{\parskip}{0pt}}

\definecolor{linky}{rgb}{0.0, 0.5, 1.0}
\newtcolorbox{repobox}{colback=red, colframe=red!75!black,
	boxrule=0.5pt, arc=2pt, left=6pt, right=6pt, top=3pt, bottom=3pt}
\newcommand{\ExternalLink}{%
	\tikz[x=1.2ex, y=1.2ex, baseline=-0.05ex]{%
		\begin{scope}[x=1ex, y=1ex]
			\clip (-0.1,-0.1)--++(-0,1.2)--++(0.6,0)--++(0,-0.6)--++(0.6,0)--++(0,-1);
			\path[draw,line width = 0.5,rounded corners=0.5](0,0) rectangle (1,1);
		\end{scope}
		\path[draw, line width = 0.5] (0.5, 0.5)--(1,1);
		\path[draw, line width = 0.5] (0.6, 1)--(1, 1)--(1, 0.6);
}}

\title{The Population Synthesis Toolkit (PST) Python Library}
\author[1]{Pablo Corcho-Caballero}
\author[2,3]{Yago Ascasibar}
\author[4]{Daniel Jim\'enez-L\'opez}
\affil[1]{Kapteyn Astronomical Institute, University of Groningen, The Netherlands}
\affil[2]{Department of Theoretical Physics, Universidad Aut\'onoma de Madrid (UAM), Spain}
\affil[3]{Centro de Investigaci\'on Avanzada en F\'isica Fundamental (CIAFF-UAM), Spain}
\affil[4]{Observatorio Astron\'omico Nacional (OAN), Spain}
\date{\vspace{-7ex}}

\begin{document}
	\maketitle
	
	\marginpar{
		
		\begin{flushleft}
			\sffamily\small
			
			{\bfseries DOI:} \href{https://doi.org/DOI unavailable}{\color{linky}{DOI unavailable}}
			
			\vspace{2mm}
			
			{\bfseries Software}
			\begin{itemize}
				\setlength\itemsep{0em}
				\item \href{N/A}{\color{linky}{Review}} \ExternalLink
				\item \href{NO_REPOSITORY}{\color{linky}{Repository}} \ExternalLink
				\item \href{DOI unavailable}{\color{linky}{Archive}} \ExternalLink
			\end{itemize}
			
			\vspace{2mm}
			
			\par\noindent\hrulefill\par
			
			\vspace{2mm}
			
			{\bfseries Editor:} \href{https://example.com}{Pending
				Editor} \ExternalLink \\
			\vspace{1mm}
			{\bfseries Reviewers:}
			\begin{itemize}
				\setlength\itemsep{0em}
				\item \href{https://github.com/Pending Reviewers}{@Pending
					Reviewers}
			\end{itemize}
			\vspace{2mm}
			
			{\bfseries Submitted:} N/A\\
			{\bfseries Published:} N/A
			
			\vspace{2mm}
			{\bfseries License}\\
			Authors of papers retain copyright and release the work under a Creative Commons Attribution 4.0 International License (\href{http://creativecommons.org/licenses/by/4.0/}{\color{linky}{CC BY 4.0}}).

		\end{flushleft}
	}
	
	\hypertarget{summary}{%
		\section{Summary}\label{summary}}
	
	Stellar population synthesis is a crucial methodology in astrophysics,
	enabling the interpretation of the integrated light of galaxies and
	stellar clusters. By combining empirical and/or theoretical libraries of
	the spectral energy distributions emitted by simple stellar populations
	(SSPs) with star formation history (SFH) and chemical evolution models,
	population synthesis can help estimate essential properties of galaxies,
	such as total stellar mass, star formation rate, mass-weighted age,
	metallicity, and so on.
	
	PST is a Python library that offers a comprehensive and flexible
	framework for stellar population synthesis. Its main goal is to easily
	and efficiently compute composite spectra using different galaxy
	evolution models and SSP libraries. It also incorporates additional
	effects such as cosmic redshift, dust extinction and attenuation, and
	computes several observable quantities derived from the spectra,
	including broadband photometric fluxes and equivalent widths.
	
	\hypertarget{state-of-the-field}{%
		\section{State of the field}\label{state-of-the-field}}
	
	A number of software packages have been developed to support stellar
	population synthesis and modeling of galaxy spectral energy
	distributions. Tools such as
	\href{https://gitlab.surrey.ac.uk/ri0005/binary_c-python}{binary\_c-python}
	(Hendriks \& Izzard, 2023) and
	\href{https://github.com/astropy/SPISEA}{SPISEA} (Hosek et al., 2020)
	are designed primarily for generating and analyzing simple stellar
	populations, often with a focus on individual stars, binaries, or star
	clusters. Meanwhile, libraries such as
	\href{https://github.com/dfm/python-fsps}{python-FSPS} (Johnson et al.,
	2024), a Python interface to the Flexible Stellar Population Synthesis
	(FSPS) code (Conroy et al., 2009; Conroy \& Gunn, 2010), and the more
	recent \href{https://github.com/ArgonneCPAC/dsps}{DSPS} (Hearin et al.,
	2023), implemented using JAX for efficient gradient computation and
	forward modeling, provide extensive modeling capabilities, although they
	are sometimes limited to a specific set of SSP models or isochrones.
	
	Other packages put a stronger emphasis on fitting observed data to
	derive galaxy properties. Examples of these include Bayesian frameworks
	such as \href{https://cigale.lam.fr/}{CIGALE} (Boquien et al., 2019),
	\href{https://github.com/asgr/ProSpect}{ProSpect} (Robotham et al.,
	2020) and \href{https://prospect.readthedocs.io/en/v1.0.0/}{Prospector}
	(Johnson et al., 2021), which infer star formation histories and other
	physical parameters using spectro-photometric data. Alternative
	frequentist tools such as \href{https://pypi.org/project/ppxf/}{PpXF}
	(Cappellari \& Emsellem, 2004),
	\href{http://www.starlight.ufsc.br/}{Starlight} (Cid Fernandes et al.,
	2005), or \href{https://gitlab.com/pipe3d/pyPipe3D}{Pipe3D} (S\'anchez et
	al., 2016), are commonly used to extract stellar kinematics and stellar
	population parameters from observed galaxy spectra, often in the context
	of integral field spectroscopy.
	
	\hypertarget{statement-of-need}{%
		\section{Statement of need}\label{statement-of-need}}
	
	The user-friendly modular framework of PST is designed to address the
	following challenges:
	
	\begin{itemize}
		\tightlist
		\item
		Handle a broad variety of SSP libraries that are publicly available in
		heterogeneous native formats.
		\item
		Model arbitrarily complex galaxy star formation and chemical evolution
		histories.
		\item
		Enable the simultaneous and self-consistent analysis of photometric
		and spectroscopic data from different instruments.
	\end{itemize}
	
	PST is designed for astronomy researchers, especially those working in
	extragalactic astrophysics and stellar population synthesis, who require
	a flexible and extensible Python-based toolkit for modeling galaxy
	properties. PST is suited for users with intermediate to advanced Python
	expertise and familiarity with common data formats and concepts in
	astronomical spectroscopy and photometry.
	
	The primary use cases are data analysis, synthetic model construction,
	and pipeline integration for studies involving stellar population
	synthesis (see the examples below). PST is particularly useful in
	workflows that combine observational data with theoretical models within
	a Bayesian or forward-modeling framework.
	
	PST is currently a dependency of
	\href{https://github.com/pykoala}{PyKOALA} (Corcho-Caballero et al.,
	2025), another open-source Python package focused on reducing optical
	integral-field spectroscopic observations. There, PST is mainly used to
	derive broadband photometry. PST is also at the core of the Bayesian
	Estimator for Stellar Population Analysis
	(\href{https://https://besta.readthedocs.io/}{BESTA}, see also Euclid
	Collaboration: Corcho-Caballero et al., 2025), where it is coupled with
	the \href{https://cosmosis.readthedocs.io/en/latest/}{CosmoSIS} (Zuntz
	et al., 2015) Monte Carlo sampling framework to infer the physical
	properties of galaxies from the observed colors and spectra.
	
	\hypertarget{features-and-functionality}{%
		\section{Features and functionality}\label{features-and-functionality}}
	
	PST design is built around three main components. First, the SSP module
	enables the consistent use and manipulation of different SSP libraries.
	This allows for the seamless ingestion of models and data from various
	literature sources. The current version includes interfaces to a range
	of SSP models, including PopStar (Moll\'a et al., 2009), Bruzual and
	Charlot (BC03, Bruzual \& Charlot, 2003), E-MILES (Vazdekis et al.,
	2016), and the X-Shooter Spectral Library (XSL, Verro et al., 2022) SSP
	models.
	
	For any SSP model integrated into PST, the library provides tools for
	interpolating across stellar ages, metallicities, and wavelengths. Users
	can easily compute key SSP quantities, such as the stellar mass-to-light
	ratio in a given band, colors, and line indices.
	
	Second, the \texttt{ChemicalEvolutionModel} classes represent the star
	formation and chemical enrichment histories required to produce
	composite spectral energy distributions and additional derived
	quantities. These classes implement several widely-used analytic
	prescriptions for modeling star formation histories (SFHs), such as
	exponentially-declining or log-normal models. They also implement
	complex SFH representations, such as table-based SFHs and particle-like
	data models. These models are particularly suitable for post-processing
	results from cosmological hydrodynamic simulations.
	
	Third, PST features a dedicated \texttt{observables} module that can
	predict additional quantities from spectra, such as broadband
	photometric fluxes, colors, and equivalent widths, which are useful for
	estimating the strength of absorption or emission lines. PST includes
	automatic integration with the photometric filters provided by the
	\href{http://svo2.cab.inta-csic.es/theory/fps/}{Spanish Virtual
		Observatory Filter Profile Service} (Rodrigo et al., 2024, 2012; Rodrigo
	\& Solano, 2020) for synthetic photometry calculations, as well as
	popular line indices such as the Lick system (Worthey et al., 1994).
	
	\hypertarget{documentation-and-tutorials}{%
		\section{Documentation and
			tutorials}\label{documentation-and-tutorials}}
	
	To make PST easier to use, we provide a set of comprehensive tutorials
	in the form of Jupyter notebooks. These tutorials cover the following
	topics:
	
	\begin{itemize}
		\tightlist
		\item
		Interacting with SSP models and exploring their fundamental
		properties.
		\item
		Producing composite spectra using:
		
		\begin{itemize}
			\tightlist
			\item
			Analytic SFH models;
			\item
			Table-based SFH models;
			\item
			Particle-like data representations.
		\end{itemize}
		\item
		Predicting observable quantities for a grid of models.
	\end{itemize}
	
	Full documentation is available
	\href{https://population-synthesis-toolkit.readthedocs.io/en/latest/}{online}.
	
	\hypertarget{acknowledgements}{%
		\section{Acknowledgements}\label{acknowledgements}}
	
	We acknowledge financial support from the Spanish State Research Agency
	(AEI/10.13039/501100011033) through grant PID2019-107408GB-C42.
	
	Daniel Jim\'enez-L\'opez was supported by Fondo Europeo de Desarrollo
	Regional (MINCIN/AEI/10.13039/501100011033/FEDER, UE), through a
	FPI-contract fellowship in the project PID2022-138560NB.
	
	Our package relies on several widely used open-source Python libraries,
	including \href{http://www.numpy.org}{NumPy} (Harris et al., 2020),
	\href{https://www.matplotlib.org/}{Matplotlib} (Hunter, 2007) and
	\href{http://www.astropy.org}{Astropy} (Astropy Collaboration et al.,
	2022, 2018, 2013).
	
	\hypertarget{references}{%
		\section*{References}\label{references}}
	\addcontentsline{toc}{section}{References}
	
	\hypertarget{refs}{}
	\begin{CSLReferences}{1}{0}
		\leavevmode\hypertarget{ref-astropy:2022}{}%
		Astropy Collaboration, Price-Whelan, A. M., Lim, P. L., Earl, N.,
		Starkman, N., Bradley, L., Shupe, D. L., Patil, A. A., Corrales, L.,
		Brasseur, C. E., N"othe, M., Donath, A., Tollerud, E., Morris, B. M.,
		Ginsburg, A., Vaher, E., Weaver, B. A., Tocknell, J., Jamieson, W.,
		\ldots{} Astropy Project Contributors. (2022). The {Astropy Project}:
		Sustaining and growing a community-oriented open-source project and the
		latest major release (v5.0) of the core package. \emph{The Astrophysical
			Journal}, \emph{935}(2), 167.
		\url{https://doi.org/10.3847/1538-4357/ac7c74}
		
		\leavevmode\hypertarget{ref-astropy:2018}{}%
		Astropy Collaboration, Price-Whelan, A. M., Sipőcz, B. M., Günther, H.
		M., Lim, P. L., Crawford, S. M., Conseil, S., Shupe, D. L., Craig, M.
		W., Dencheva, N., Ginsburg, A., Vand erPlas, J. T., Bradley, L. D.,
		P\'erez-Su\'arez, D., de Val-Borro, M., Aldcroft, T. L., Cruz, K. L.,
		Robitaille, T. P., Tollerud, E. J., \ldots{} Astropy Contributors.
		(2018). The {Astropy Project}: Building an open-science project and
		status of the v2.0 core package. \emph{The Astronomical Journal},
		\emph{156}(3), 123. \url{https://doi.org/10.3847/1538-3881/aabc4f}
		
		\leavevmode\hypertarget{ref-astropy:2013}{}%
		Astropy Collaboration, Robitaille, T. P., Tollerud, E. J., Greenfield,
		P., Droettboom, M., Bray, E., Aldcroft, T., Davis, M., Ginsburg, A.,
		Price-Whelan, A. M., Kerzendorf, W. E., Conley, A., Crighton, N.,
		Barbary, K., Muna, D., Ferguson, H., Grollier, F., Parikh, M. M., Nair,
		P. H., \ldots{} Streicher, O. (2013). Astropy: A community {Python}
		package for astronomy. \emph{Astronomy \& Astrophysics}, \emph{558},
		A33. \url{https://doi.org/10.1051/0004-6361/201322068}
		
		\leavevmode\hypertarget{ref-boquien+19}{}%
		Boquien, M., Burgarella, D., Roehlly, Y., Buat, V., Ciesla, L., Corre,
		D., Inoue, A. K., \& Salas, H. (2019). {CIGALE: a python Code
			Investigating GALaxy Emission}. \emph{Astronomy \& Astrophysics},
		\emph{622}, A103. \url{https://doi.org/10.1051/0004-6361/201834156}
		
		\leavevmode\hypertarget{ref-bc+03}{}%
		Bruzual, G., \& Charlot, S. (2003). {Stellar population synthesis at the
			resolution of 2003}. \emph{Monthly Notices of the Royal Astronomical
			Society}, \emph{344}(4), 1000--1028.
		\url{https://doi.org/10.1046/j.1365-8711.2003.06897.x}
		
		\leavevmode\hypertarget{ref-capellari+04}{}%
		Cappellari, M., \& Emsellem, E. (2004). Parametric recovery of
		line-of-sight velocity distributions from absorption-line spectra of
		galaxies via penalized likelihood. \emph{Publications of the
			Astronomical Society of the Pacific}, \emph{116}(816), 138--147.
		\url{https://doi.org/10.1086/381875}
		
		\leavevmode\hypertarget{ref-cid-fernandes+05}{}%
		Cid Fernandes, R., Mateus, A., Sodr\'e, L., Stasi\'nska, G., \& Gomes, J. M.
		(2005). {Semi-empirical analysis of Sloan Digital Sky Survey galaxies -
			I. Spectral synthesis method}. \emph{Monthly Notices of the Royal
			Astronomical Society}, \emph{358}(2), 363--378.
		\url{https://doi.org/10.1111/j.1365-2966.2005.08752.x}
		
		\leavevmode\hypertarget{ref-conroyux5cux26gunn10}{}%
		Conroy, C., \& Gunn, J. E. (2010). The propagation of uncertainties in
		stellar population synthesis modeling. {III}. Model calibration,
		comparison, and evaluation. \emph{The Astrophysical Journal},
		\emph{712}(2), 833--857.
		\url{https://doi.org/10.1088/0004-637X/712/2/833}
		
		\leavevmode\hypertarget{ref-conroy+09}{}%
		Conroy, C., Gunn, J. E., \& White, M. (2009). The propagation of
		uncertainties in stellar population synthesis modeling. {I}. The
		relevance of uncertain aspects of stellar evolution and the initial mass
		function to the derived physical properties of galaxies. \emph{The
			Astrophysical Journal}, \emph{699}(1), 486--506.
		\url{https://doi.org/10.1088/0004-637X/699/1/486}
		
		\leavevmode\hypertarget{ref-pykoala_adass}{}%
		Corcho-Caballero, P., Ascasibar, Y., L\'opez-S\'anchez, A. R.,
		Gonz\'alez-Bol\'ivar, M., Lorente, N., Tocknell, J., Jim\'enez-Ibarra, F.,
		Daluwathumullagamage, P. J., Quattropani, G., Owers, M., \&
		Verdoes-Kleijn, G. A. (2025). \emph{The {PyKOALA} {Python} library: A
			multi-instrument package for {IFS} data reduction}.
		\url{https://arxiv.org/abs/2507.18347}
		
		\leavevmode\hypertarget{ref-cc+25}{}%
		Euclid Collaboration: Corcho-Caballero, P., Ascasibar, Y., Verdoes
		Kleijn, G., Lovell, C. C., De Lucia, G., Cleland, C., Fontanot, F.,
		Tortora, C., Koopmans, L. V. E., Moutard, T., Laigle, C., Nersesian, A.,
		Shankar, F., Aghanim, N., Altieri, B., Amara, A., Andreon, S., Aussel,
		H., Baccigalupi, C., \ldots{} Tenti, M. (2025). {Euclid Quick Data
			Release (Q1). A probabilistic classification of quenched galaxies}.
		\emph{arXiv e-Prints}, arXiv:2503.15315.
		\url{https://doi.org/10.48550/arXiv.2503.15315}
		
		\leavevmode\hypertarget{ref-harris2020array}{}%
		Harris, C. R., Millman, K. J., Walt, S. J. van der, Gommers, R.,
		Virtanen, P., Cournapeau, D., Wieser, E., Taylor, J., Berg, S., Smith,
		N. J., Kern, R., Picus, M., Hoyer, S., Kerkwijk, M. H. van, Brett, M.,
		Haldane, A., R\'io, J. F. del, Wiebe, M., Peterson, P., \ldots{} Oliphant,
		T. E. (2020). Array programming with {NumPy}. \emph{Nature},
		\emph{585}(7825), 357--362.
		\url{https://doi.org/10.1038/s41586-020-2649-2}
		
		\leavevmode\hypertarget{ref-hearin+23}{}%
		Hearin, A. P., Chaves-Montero, J., Alarcon, A., Becker, M. R., \&
		Benson, A. (2023). {DSPS: Differentiable stellar population synthesis}.
		\emph{Monthly Notices of the Royal Astronomical Society}, \emph{521}(2),
		1741--1756. \url{https://doi.org/10.1093/mnras/stad456}
		
		\leavevmode\hypertarget{ref-hendriksux5cux26Izzard23}{}%
		Hendriks, D., \& Izzard, R. (2023). {binary\_c-python: A Python-based
			stellar population synthesis tool and interface to binary\_c}. \emph{The
			Journal of Open Source Software}, \emph{8}(85), 4642.
		\url{https://doi.org/10.21105/joss.04642}
		
		\leavevmode\hypertarget{ref-hosek+20}{}%
		Hosek, M. W., Jr., Lu, J. R., Lam, C. Y., Gautam, A. K., Lockhart, K.
		E., Kim, D., \& Jia, S. (2020). {SPISEA}: A {Python}-based simple
		stellar population synthesis code for star clusters. \emph{The
			Astronomical Journal}, \emph{160}(3), 143.
		\url{https://doi.org/10.3847/1538-3881/aba533}
		
		\leavevmode\hypertarget{ref-hunter:2007}{}%
		Hunter, J. D. (2007). Matplotlib: A {2D} graphics environment.
		\emph{Computing in Science \& Engineering}, \emph{9}(3), 90--95.
		\url{https://doi.org/10.1109/MCSE.2007.55}
		
		\leavevmode\hypertarget{ref-jonhson+24}{}%
		Johnson, B. D., Foreman-Mackey, D., Sick, J., Leja, J., Walmsley, M.,
		Tollerud, E., Leung, H., Scott, S., \& Park, M. (2024).
		\emph{{dfm/python-fsps: v0.4.7}} (Version v0.4.7) {[}Computer
		software{]}. Zenodo. \url{https://doi.org/10.5281/zenodo.12447779}
		
		\leavevmode\hypertarget{ref-johnson+21}{}%
		Johnson, B. D., Leja, J., Conroy, C., \& Speagle, J. S. (2021). Stellar
		population inference with {Prospector}. \emph{The Astrophysical Journal
			Supplement Series}, \emph{254}(2), 22.
		\url{https://doi.org/10.3847/1538-4365/abef67}
		
		\leavevmode\hypertarget{ref-molla+09}{}%
		Moll\'a, M., Garcı\'ia-Vargas, M. L., \& Bressan, A. (2009). {PopStar I}:
		Evolutionary synthesis model description. \emph{Monthly Notices of the
			Royal Astronomical Society}, \emph{398}(1), 451--470.
		\url{https://doi.org/10.1111/j.1365-2966.2009.15160.x}
		
		\leavevmode\hypertarget{ref-robotham+20}{}%
		Robotham, A. S. G., Bellstedt, S., Lagos, C. del P., Thorne, J. E.,
		Davies, L. J., Driver, S. P., \& Bravo, M. (2020). {ProSpect}:
		Generating spectral energy distributions with complex star formation and
		metallicity histories. \emph{Monthly Notices of the Royal Astronomical
			Society}, \emph{495}(1), 905--931.
		\url{https://doi.org/10.1093/mnras/staa1116}
		
		\leavevmode\hypertarget{ref-rodrigo+24}{}%
		Rodrigo, C., Cruz, P., Aguilar, J. F., Aller, A., Solano, E.,
		G\'alvez-Ortiz, M. C., Jim\'enez-Esteban, F., Mas-Buitrago, P., Bayo, A.,
		Cort\'es-Contreras, M., Murillo-Ojeda, R., Bonoli, S., Cenarro, J., Dupke,
		R., L\'opez-Sanjuan, C., Mar\'in-Franch, A., de Oliveira, C. M., Moles, M.,
		Taylor, K., \ldots{} Rami\'o, H. V. (2024). {Photometric segregation of
			dwarf and giant FGK stars using the SVO Filter Profile Service and
			photometric tools}. \emph{Astronomy \& Astrophysics}, \emph{689}, A93.
		\url{https://doi.org/10.1051/0004-6361/202449998}
		
		\leavevmode\hypertarget{ref-rodrigo+20}{}%
		Rodrigo, C., \& Solano, E. (2020). {The SVO Filter Profile Service}.
		\emph{XIV.0 Scientific Meeting (Virtual) of the Spanish Astronomical
			Society}, 182.
		
		\leavevmode\hypertarget{ref-rodrigo+12}{}%
		Rodrigo, C., Solano, E., \& Bayo, A. (2012). \emph{{SVO Filter Profile
				Service Version 1.0}} (p. 1015). IVOA Working Draft 15 October 2012.
		\url{https://doi.org/10.5479/ADS/bib/2012ivoa.rept.1015R}
		
		\leavevmode\hypertarget{ref-sanchez+16}{}%
		S\'anchez, S. F., P\'erez, E., S\'anchez-Bl\'azquez, P., Gonz\'alez, J. J.,
		Ros\'ales-Ortega, F. F., Cano-D\'iaz, M., L\'opez-Cob\'a, C., Marino, R. A., Gil
		de Paz, A., Moll\'a, M., L\'opez-S\'anchez, A. R., Ascasibar, Y., \&
		Barrera-Ballesteros, J. (2016). {Pipe3D}, a pipeline to analyze integral
		field spectroscopy data: {I}. New fitting philosophy of {FIT3D}.
		\emph{Revista Mexicana de Astronom\'ia y Astrof\'isica}, \emph{52}, 21--53.
		\url{https://doi.org/10.48550/arXiv.1509.08552}
		
		\leavevmode\hypertarget{ref-vazdekis+16}{}%
		Vazdekis, A., Koleva, M., Ricciardelli, E., Röck, B., \& Falc\'on-Barroso,
		J. (2016). {UV}-extended {E-MILES} stellar population models: Young
		components in massive early-type galaxies. \emph{Monthly Notices of the
			Royal Astronomical Society}, \emph{463}(4), 3409--3436.
		\url{https://doi.org/10.1093/mnras/stw2231}
		
		\leavevmode\hypertarget{ref-verro+22}{}%
		Verro, K., Trager, S. C., Peletier, R. F., Lançon, A., Arentsen, A.,
		Chen, Y.-P., Coelho, P. R. T., Dries, M., Falc\'on-Barroso, J., Gonneau,
		A., Lyubenova, M., Martins, L., Prugniel, P., S\'anchez-Bl\'azquez, P., \&
		Vazdekis, A. (2022). {Modelling simple stellar populations in the
			near-ultraviolet to near-infrared with the X-shooter Spectral Library
			(XSL)}. \emph{Astronomy \& Astrophysics}, \emph{661}, A50.
		\url{https://doi.org/10.1051/0004-6361/202142387}
		
		\leavevmode\hypertarget{ref-worthey+94}{}%
		Worthey, G., Faber, S. M., Gonzalez, J. J., \& Burstein, D. (1994). Old
		stellar populations. {V}. Absorption feature indices for the complete
		{Lick/IDS} sample of stars. \emph{The Astrophysical Journal Supplement
			Series}, \emph{94}, 687. \url{https://doi.org/10.1086/192087}
		
		\leavevmode\hypertarget{ref-zuntz+15}{}%
		Zuntz, J., Paterno, M., Jennings, E., Rudd, D., Manzotti, A., Dodelson,
		S., Bridle, S., Sehrish, S., \& Kowalkowski, J. (2015). {CosmoSIS}:
		Modular cosmological parameter estimation. \emph{Astronomy and
			Computing}, \emph{12}, 45--59.
		\url{https://doi.org/10.1016/j.ascom.2015.05.005}
		
	\end{CSLReferences}
	
\end{document}